\magnification=\magstep1
\tolerance=10000
\centerline{\bf PATTERNS OF LORENTZ SYMMETRY BREAKING}
\centerline{\bf IN QED BY  CPT-ODD INTERACTION} 

\bigskip

\centerline{{\bf A. A. Andrianov}
\footnote{$^\natural$}{ E-mail:  andrian@snoopy.phys.spbu.ru}}
\centerline{\sl Department of Theoretical Physics,}
\centerline{\sl Sankt-Petersburg State University,}
\centerline{\sl 198904 Sankt-Petersburg, Russia}
\centerline{ and}
\centerline{\sl  Departament d'Estructura i Constituents de la Mat\`eria}
\centerline{\sl Universitat de Barcelona }
\centerline{\sl  Diagonal, 647, 08028 Barcelona, Spain}
\medskip

\centerline{{\bf R. Soldati}
\footnote{$^\flat$}{ E-mail: soldati@bo.infn.it}}
\centerline{\sl Dipartimento di Fisica "A. Righi", Universit\'a
di Bologna and Istituto Nazionale}
\centerline{\sl di Fisica Nucleare, Sezione di Bologna,
40126 Bologna, Italia}

\vskip .5cm

\baselineskip=10 pt

\centerline{\bf Abstract}

\bigskip

{\it A tiny Lorentz symmetry breaking can be mediated in Electrodynamics
by means 
of the Chern-Simons (CS)
interaction polarized along a constant CS vector. Its presence makes the
vacuum optically active that has been recently estimated from
astrophysical data. We examine two possibilities for the CS vector to 
be time-like or 
space-like, under the assumption that it originates from v.e.v. 
of some pseudoscalar matter. It is shown that: \quad
a) a time-like CS vector makes the vacuum unstable under pairs creation
of tachyonic photon modes with the finite vacuum decay rate, i.e.
it is unlikely realized at macroscopic time scales; \quad
b) on the contrary, the space-like CS vector does not yield any tachyonic
modes and, moreover, if its dynamical counterpart is substantially described
by a scale invariant interaction, then the QED radiation effects
induce the dynamical breaking of Lorentz symmetry, i.e.
the occurrence of space-like CS vector appears to be rather natural.}

\vfill
\bigskip
\noindent 
DFUB-98/8\quad UB-ECM-PF-98/10\quad hep-ph/9804448
\par\noindent
April 98
\medskip
\centerline{\sl Submitted to Phys. Lett. B}
\eject

\baselineskip=22 pt

\medskip

The possibility of a tiny deviation from the 
Lorentz covariance  has
been recently reconsidered [1-5] and severe 
astrophysical bounds  on it have been
derived. In Electrodynamics,
when one retains its fundamental
character provided by the renormalizability, it is 
conceivable to have two types
of the Lorentz symmetry breaking (LSB). One way is for the light
velocity to depart from the maximum attainable speed for other particles [3].
Another way is under the scope of our letter and is delivered in the
ordinary {\it 3+1} dimensional Minkowski space-time by
the (C)PT-odd Chern-Simons (CS) coupling of photons to 
the vacuum [2] mediated
by a constant CS vector $\eta_\mu$ (Carroll-Field-Jackiw model):
$$
{\cal L}_{{\rm QED}} =
-{1\over 4}F_{\mu\nu}F^{\mu\nu} + 4\pi A_\mu J^\mu
+ {1\over 2}\epsilon^{\mu\nu\lambda\sigma} A_\mu F_{\nu\lambda} \eta_\sigma ,
\eqno(1)
$$
where we have included also the photon coupling to a conserved matter current
$J^\mu$. This modification of Electrodynamics does not break the 
gauge symmetry of the action but splits the dispersion relations 
for different photon (circular) polarizations [2].
As a consequence the {\it linearly} polarized photons exhibit 
the birefringence, {\it i.e.} the
rotation of the polarization direction depending on the distance, when they
propagate in the vacuum.

If the vector $\eta_\mu$ is time-like, $\eta^2 > 0$, then 
this observable effect may be isotropic in the preferred frame
(say, the Rest Frame of the Universe where the cosmic microwave background
radiation is maximally isotropic), since $\eta_\mu = (\eta_0,0,0,0)$.
However it is definitely anisotropic
for space-like  $\eta^2 < 0 $. The first possibility was carefully
analyzed [2,5] resulting in the bound:
$|\eta_0| < 10^{-33} eV \simeq 10^{-28} cm^{-1}$. Last year Nodland and Ralston
[4] presented their compilation of data on polarization rotation of
photons from remote radio galaxies and argued for the existence of
space anisotropy with  $\eta_\mu \simeq (0, \vec\eta)$ of order
$|\vec\eta| \sim 10^{-32} eV \simeq 10^{-27} cm^{-1}$. 
Thereupon the hot discussion started about the confidence level 
of their result [6] and still this effect needs a better confirmation [7].

In our letter we would like to draw the reader's attention to the fact that
the time-like pattern for the CS interaction is intrinsically inconsistent
as it is accompanied by the creation of tachyonic photon modes
\footnote{$^1$}{The presence of tachyonic modes in the photon spectrum
and the posssible vacuum instability was mentioned in [2].}
from the vacuum, {\it i.e.} such a vacuum is unstable under the QED radiative
effects. On the contrary, the space-like anisotropy carrying CS vector 
does not generate any vacuum instability 
and may be naturally induced [8]
by a Coleman-Weinberg mechanism [9] in any scale invariant scenario where
the CS vector is related to v.e.v. of the gradient of a pseudoscalar
field \footnote{$^2$}{ It may be also interpreted as 
the average value of a background torsion field [10] with similar 
conclusions.}.

Let us give more details concerning the above statements. 
The CS interaction modifies the second couple of the Maxwell equations:
$$
\partial_\mu  F^{\mu\nu} = 4\pi J^\nu + 2 \eta_\mu \widetilde  F^{\mu\nu},
\eqno(2)
$$
where the notation $\widetilde  F^{\mu\nu} \equiv 
{1\over 2}\epsilon^{\mu\nu\lambda\sigma} F_{\lambda\sigma}$
is adopted. One can read off this equation that for $\vec\eta \not= 0$ 
the electric field strength $\vec E$ 
is no longer orthogonal to the wave vector because 
$\nabla\cdot\vec E \not= 0$ in the empty space. Thus, in this case, 
the photon obtains a third polarization like a massive vector particle.

The photon energy spectrum can be derived from 
the wave equations on the gauge
potential $A_\mu$ in the momentum
representation:
$$
\eqalign{
\left( p^2 g^{\mu\nu} + 2 i \epsilon^{\mu\nu\lambda\sigma} 
\eta_\lambda p_\sigma
\right) A_\nu & \equiv {\rm \bf K}^{\mu\nu}  A_\nu = 0;\cr
p^\mu  A_\mu & = 0. \cr}
\eqno(3)
$$
It is evident that the CS interaction changes the  photon
spectrum only in the polarization hyper-plane orthogonal to the momentum
$p_\mu$ and the CS vector $\eta_\nu$. The projector on this plane has the form:
$$
{\rm\bf e}^{\mu\nu}_2 =  g^{\mu\nu}  - {(\eta^\mu p^\nu + \eta^\nu p^\mu)
(\eta\cdot p) - \eta^\mu \eta^\nu p^2 - p^\mu p^\nu \eta^2 \over 
(\eta\cdot p)^2 - \eta^2 p^2}.
\eqno(4)
$$
After introducing the notation,
$$
{\cal E}^{\mu\nu} \equiv 2 i \epsilon^{\mu\nu\lambda\sigma} 
\eta_\lambda p_\sigma,
\eqno(5)
$$
one can verify that\footnote{$^3$}{ In 
what follows the matrix product  
 is provided by  contraction with  $g_{\mu\nu}$.}
$$
{\rm\bf e}_2 = 
{{\cal E} \cdot {\cal E} \over {\rm N}};
\qquad {\rm N} \equiv 4 ((\eta\cdot p)^2 - \eta^2 p^2),
\eqno(6)
$$
and $ {\cal E}\cdot {\rm\bf e}_2 = {\cal E}$. Respectively,
 one can unravel
the energy spectrum of the wave equation (3) in terms of two
polarizations of different helicity:
$$
{\rm\bf e}_{L, R} = {1 \over 2} \left({\rm\bf e}_2 \pm
{{\cal E} \over \sqrt{{\rm N}}} \right);\qquad
{\cal E} \cdot {\rm\bf e}_{L, R} = \pm \sqrt{{\rm N}} {\rm\bf e}_{L, R}.
\eqno(7)
$$
Then the dispersion relation is controlled by the equation:
$$
(p^2)^2 + 4 \eta^2 p^2 - 4 (\eta\cdot p)^2 = 0.
\eqno(8)
$$
From eq.~(8) one discovers the different physical properties depending
on whether $\eta^2$ is positive or negative.

If $\eta^2 > 0$ one can examine photon properties in the rest frame for
the CS vector  $\eta_\mu = (\eta_0,0,0,0)$. Then the dispersion relation,
$$
(p_0)_\pm^2 = \vec p^2 \pm 2 |\eta_0| |\vec p|,
\eqno(9)
$$
tells us that the upper type of solutions can be interpreted as 
describing massless states because
their energies vanish
 for $\vec p =0$ .
Meanwhile  the lower type of distorted photons is such that they
behave as tachyons [2] 
 with a real energy for $|\vec p| >  2 |\eta_0|$
(when their phase 
velocity are taken into account) . 
There are also static solutions with $p_0 =0 \Leftrightarrow 
|\vec p| = 2 |\eta_0|$ and 
unstable solutions (tachyons) with 
a negative imaginary energy for $|\vec p| <  2 |\eta_0|$.

For light-like  CS vectors, $\eta^2 = 0$, one deals with  conventional photons
of shifted energy-momentum spectra for different polarizations:
$$
(p_0 \pm \eta_0 )^2 =(\vec p \pm \vec\eta )^2 .
\eqno(10) 
$$

If the CS vector is space-like, $\eta^2 < 0$, the photon spectrum is 
more transparent
in the static frame where $\eta_\mu = (0, \vec\eta)$. The corresponding
dispersion relation reads:
$$
\left(p_0\right)_{\pm}^2 = 
\vec p^2 + 2\vec\eta^2 \pm 2 \sqrt{|\vec\eta|^4 + (\vec\eta \cdot \vec p)^2}.
\eqno(11)
$$
It can be checked that in this case $p_0^2 \geq 0$ for all $\vec p$ 
and neither static nor unstable tachyonic modes do actually arise. 
The upper type of solution
describes the massive particle with a mass $m_+ = 2  |\vec\eta|$  
for small space momenta $|\vec p| \ll |\vec\eta|$. The lower type of solutions
represents a massless state as $p_0 \rightarrow 0 $ for $|\vec p| 
\rightarrow 0$. It might also exhibit the acausal behavior 
when $p_\mu  p^\mu < 0$ but, even
in this case $p_0^2 \geq 0$ for all $\vec p$, so that 
the unstable tachyonic modes never arise.

In a general frame, for high momenta $|\vec p|\gg |\vec\eta|$, 
$|p_0|\gg \eta_0$, 
one obtains the relation:
$$
|p_0| - |\vec p| \simeq \pm (\eta_0 -|\vec\eta| \cos\varphi ),
\eqno(12)
$$ 
where $\varphi$ is an angle between $\vec\eta$ and $\vec p$.
Hence, for a given photon frequency $p_0$, the 
phase shift induced by the difference between  wave vectors
of opposite helicities does not depend upon this frequency. Moreover
the linearly polarized waves - a combination of left- and right-handed ones -
 reveal the birefringence phenomenon of rotation of polarization axis with
the distance [2].

Let us now examine the radiative effects induced by the emission of
distorted photons. 
In principle the energy and momentum 
conservation allows for pairs
of tachyons to be created from the vacuum due to the CS interaction.
Thereby in any model where the CS vector plays a dynamical role, being related
to the condensate of a matter field, one may
expect that, owing to tachyon pairs creation,
the asymptotic Fock's vacuum state becomes unstable and 
transforming towards a true non-perturbative state without 
tachyonic photon modes. But if we inherit the causal prescription for 
propagating physical waves, then the physical states are assigned to possess
the non-negative energy sign. As a consequence 
the tachyon pairs 
can be created out of the vacuum only if $\eta^2 > 0$. In particular, the static
waves with $p_0 =0\Leftrightarrow |\vec p| = 2 |\eta_0|$ 
are well produced to destroy the
vacuum state. On the contrary, for $\eta^2 < 0$ the causal prescription for
the energy sign together with the energy-momentum conservation prevent
the vacuum state from photon pair emission.

Thus the decay process holds when static and unstable
tachyonic modes exist. Let us  clarify this point
with the help of the radiatively induced effective potential for the
variable $\eta_\mu$ treated as an average 
value
\footnote{$^4$}{
It may be a mean value over  large
volume for slowly varying classical background field or, 
eventually, the v.e.v. for an axion-type field.}  
of the gradient of a pseudoscalar
field, $ \eta_\mu = \langle\partial_\mu\Theta(x)\rangle_\Omega$, \quad $\Omega$
being the non-perturbative vacuum state.
We adopt a model with classically scale-invariant action which naturally
extends the CFJ model~(1):
$$
S = S_{\rm Maxwell} + \int d^4x \left({1\over 2}\epsilon^{\mu\nu\lambda\sigma} 
A_\mu F_{\nu\lambda} \partial_\sigma\Theta + 
c_1  (\partial_\mu\Theta \partial^\mu\Theta)^2 + 
c_2 (\partial^2\Theta)^2\right)\ ,
\eqno(13)
$$
wher $c_1$ and $c_2$ are coefficients to be 
suitably tuned in terms of ultraviolet
renormalization effects.
The effective potential for slowly varying $\partial_\mu\Theta(x) 
\simeq  \eta_\mu$ - due to quantum photon fluctuations - 
can obtained after integration
over the photon field from the functional determinant 
(see [8] for more details):
$$
{\cal V}_{\rm ef\/f} (s^2) = c_1 (\eta_\mu \eta^\mu)^2 +  
{1 \over 4} ({\rm vol})_4^{-1} 
{\tt Tr }\left[{\rm\bf e}_2
\ln\left({{\rm \bf K}{\rm \bf K}^\top\over \mu^4}\right)\right]_{\rm reg},
\eqno(14)
$$
where $\mu$ is some infrared normalization scale, 
whilst ${\rm \bf K}^\top$ stands for the transposed matrix ${\rm \bf K}$
and the finite part of the last functional is derived with the help of
any regularization which respects the classical scale invariance.
However the very latter invariance turns out [8] to be broken 
when infrared divergences are cured. It finally yields the following 
expression for the  effective
potential:
$$
{\cal V}_{{\rm ef\/f}} = {5 \over 32\pi^2} (\eta_\mu \eta^\mu)^2 \ln\left(
{- \eta_\mu \eta^\mu \over \mu^2 \sqrt{e}}\right),
\eqno(15)
$$
where the constant $\sqrt e \ ( \ln\sqrt e =1/2) $ is factorized out  
for a further convenience and 
the infrared normalization scale $\mu$ has 
to be of the order of $10^{-32} eV$ in order
to fit the Nodland-Ralston effect [4]. 
One can see from 
eq.~(15) that: 

a) if $\eta^2 > 0$ there appears an imaginary part for the vacuum energy,
$$
{\rm Im}\ {\cal V}_{{\rm ef\/f}} = - {5 \over 32\pi} (\eta_\mu \eta^\mu)^2,
\eqno(16)
$$
which characterizes the rate per unit volume
of tachyon pairs production out of the vacuum state;

b) for  $\eta^2 \leq 0$ the effective potential is real and has a maximum
at  $\eta^2 = 0$,  whereas
the true minima arise at non-zero space-like value 
$\eta^2 = - \mu^2 $, just realizing
the celebrated Coleman-Weinberg phenomenon [9].

We conclude that it is unlikely to have the Lorentz symmetry breaking
by the CPT odd interaction (1) by means of a time-like CS vector 
preserving the rotational invariance in the $\eta_\mu$ rest frame.
Rather intrinsically, the pseudoscalar matter interacting with photons has a 
tendency to condensate along a space-like direction. In turn, as we have seen,
it leads to the photon mass formation. Of course, this effect of a 
Coleman-Weinberg type
does not yield any explanation for the  magnitude of the scale $\mu$,
which, however, is  implied to be a physical infrared cutoff 
of a cosmological origin.
Therefore its magnitude can be thought to be the inverse of the maximal 
photon wave length in the Universe: namely, $\lambda_{{\rm max}}
= 1/\mu \simeq 10^{27}cm$.

If an explicit breaking of the scale invariance is present, then
the LSB effect may be reduced and even 
disappear. For instance [8],
with the  positive kinetic term for the $\Theta$ field, 
$$
 {\cal L}_{\rm kin} = {M^2 \over 2} \partial_\nu\Theta \partial^\nu\Theta
\eqno(17)
$$
added to (13),
one can derive the maximal scale $M_{{\rm max}}$ for which
the LSB minima exist and are at their face value: 
$$
 M_{\rm max} = {\sqrt{5}\over 4\pi} \mu\ ;\qquad |\eta_\mu|_{\rm min}
= {\mu e^{- 1/4}}, 
\eqno(18)
$$
{\it i.e.} $M_{\rm max}$ of the order of $ 10^{-33}eV$.

In this letter we do not touch upon the origin of the pseudoscalar field
[10, 11], but leave this discussion to a forthcoming paper. 

\medskip 

\noindent
{\bf Acknowledgements.}
\medskip We thank R. Jackiw for useful and timely remarks.
A.A. is grateful to R. Tarrach for stimulating discussions.
This work is supported by Italian grant MURST-quota 40\%;
A.A. is also supported by grants RFFI 98-02-18137, GRACENAS 6-19-97
and  by  Spanish Ministerio de Educaci\'on y Cultura.

\bigskip
\noindent
{\bf References}
\medskip
\item{[1]}\ H. B. Nielsen and I. Picek, Nucl. Phys. {\bf B211} (1983) 269.
\item{[2]}\ S. M. Carrol, G. B. Field and R. Jackiw, Phys. Rev. {\bf D41} 
            (1990) 1231.
\item{[3]}\  S. Coleman and S. L. Glashow, Phys. Lett. {\bf 405B} (1997) 249.
\item{[4]}\  B. Nodland and J. P. Ralston, Phys. Rev. Lett. {\bf 78}
             (1997) 3043;
             \quad {\it ibid.} {\bf 79} (1997) 1958.
\item{[5]}\  D. Harari and P. Sikivie, Phys. Lett. {\bf 289B} (1992) 67;  
\item{   }\  S. Mohanty and S. N. Nayak, Phys. Rev. {\bf D48} (1993) 1526;
\item{   }\  A. Cimatti, S. di Serego Alighieri, G. B. Field and
            R. A. E. Fosbury, Astrophys. J. {\bf 422} (1994) 562;
\item{   }\  M. P. Haugan and T. F. Kauffmann, Phys. Rev. {\bf D52} (1995) 3168;
\item{   }\  M. Goldhaber and V. Trimble, J. Astrophys. Astr. {\bf 17} 
         (1996) 17.
\item{[6]}\ D. J. Eisenstein and E. F. Bunn, Phys. Rev. Lett. {\bf 79}
           (1997) 1957;
\item{   }\ S. M. Carrol and G. B. Field,  Phys. Rev. Lett. {\bf 79}
           (1997) 2934; 
\item{   }\  J. P. Leahy, {\tt astro-ph/9704285};
\item{   }\ J. F. C. Wardle, R. A. Perley and M. H. Cohen,  
     Phys. Rev. Lett. {\bf 79} (1997) 1801;
\item{   }\ T. J. Loredo, E. E. Flanagan and I. M. Wasserman, 
    Phys. Rev. {\bf D56} (1997) 7057;
\item{   }\  B. Nodland and J. P. Ralston, {\tt astro-ph/9706126,
      astro-ph/9708114}.
\item{[7]}\ P. Jain and J. P. Ralston,  {\tt astro-ph/9803164}.
\item{[8]}\ A. A. Andrianov and R. Soldati, Phys. Rev. {\bf D51} (1995) 5961;
            \ Proc. 11th Int. Workshop QFTHEP96 (St.Petersburg, 1996) 
            (MSU Publ., Moscow, 1997) 290  ({\tt hep-th/9612156}).
\item{[9]}\ S. Coleman and E. Weinberg, Phys. Rev. {\bf D7} (1973) 1888.
\item{[10]}\ A. Dobado and A. L. Maroto, Phys. Rev. {\bf D54} (1996) 5185;\quad
           Mod. Phys. Lett. {\bf A12} (1997) 3003. 
\item{[11]}\ W.-T. Ni, Phys. Rev. Lett. {\bf 38} (1977) 301;
\item{    }\ V. De Sabbata and M. Gasperini, Phys. Lett. {\bf 83A} (1981) 115;
\item{    }\ D. V. Ahluwalia and T. Goldman, Mod. Phys. Lett. {\bf A28}
             (1993) 2623;
\item{    }\ T. Banks and M. Dine, Nucl. Phys. {\bf B505} (1997) 445. 

\vfill\eject\end